\documentclass[twocolumn,10pt]{article} 

\usepackage[square,numbers,sort&compress,comma]{natbib}

\usepackage{amsmath}
\usepackage{amssymb}
\usepackage{caption}
\usepackage{graphicx}
\usepackage{latexsym}
\usepackage{times}
\usepackage{subfigure}
\usepackage{float}  
\usepackage{lipsum}  

\topmargin - 12pt 
\renewenvironment{abstract}%
              {
               \small
               {\bfseries \abstractname}
               \par
               \vspace{10pt}
              }

\renewcommand\abstractname{Abstract}

\newcommand{\nomenclature}
              [1]
              {
               \bgroup
               \flushleft
               \small\bf
               #1
               \par
               \egroup
              }

\renewcommand{\section}
              [1]
              {
               \bgroup
               \flushleft
               \small\bf
               \stepcounter{section}
               \arabic{section}. #1
               \par
               \egroup
              }

\renewcommand{\subsection}
              [1]
              {
               \bgroup
               \flushleft
               \small\em
               \stepcounter{subsection}
               \arabic{section}.
               \arabic{subsection}. #1
               \par
               \egroup
              }

\renewcommand{\subsubsection}
              [1]
              {
               \bgroup
               \flushleft
               \small\em
               \stepcounter{subsubsection}
               \arabic{section}.
               \arabic{subsection}.
               \arabic{subsubsection}. #1
               \par
               \egroup
              }

  \newcommand{\acknowledgement}
              [1]
              {
               \bgroup
               \flushleft
               \small\bf
               #1
               \par
               \egroup
              }

  \newcommand{\sectionbib}
              [1]
              {
               \bgroup
               \flushleft
               \small\bf
               #1
               \par
               \egroup
              }

\newcommand{\chem}[1]{\ensuremath{\mathrm{#1}}}

\newcommand{{\asfs}}{\ensuremath{\alpha_\mathrm{sfs}}}
\newcommand{{\lci}}{\ensuremath{\lambda_\mathrm{ci}}}

\begin{document}

\title{\LARGE Influence of flow structures and heat release on cross-scale turbulent kinetic energy transfer in premixed swirl flames.}

\author{{\large Askar Kazbekov, Adam M. Steinberg}\\[10pt]
        {\footnotesize \em $^a$Daniel Guggenheim School of Aerospace Engineering, Georgia Institute of Technology, Atlanta 30332, USA}\\[-5pt]}

\date{}


\small
\baselineskip 10pt


\twocolumn[\begin{@twocolumnfalse}
\vspace{50pt}
\maketitle
\vspace{40pt}
\rule{\textwidth}{0.5pt}
\begin{abstract}

\end{abstract}
This paper experimentally analyzes the simultaneous influence of heat release and coherent vortex structures on the transfer of kinetic energy across scales around the laminar flame thickness $\delta_\mathrm{L}^0$ in a turbulent premixed swirl flame and a non-reacting swirl flow. High-resolution tomographic particle image velocimetry and formaldehyde planar laser induced fluorescence measurements are used to obtain 3D velocity fields and estimates of the progress variable and density fields. The kinetic energy transfer across a filter scale of $\Delta = 1.5\delta_\mathrm{L}^0$ was then quantified using physical space analysis. Coherent flow structures were identified using the swirling strength and proper orthogonal decomposition was used to identify the dominant periodic flow structure. While non-reacting regions of the flow show mean down-scale energy transfer (forward-scatter), mean back-scatter is observed internal to the flame. Importantly, the back-scatter magnitude in the flame increases in regions undergoing flame/vortex interaction. That is, the mean back-scatter magnitude at locations simultaneously inside the flame and a large-scale coherent vortex is higher than regions in the flame and not in a vortex, with the mean back-scatter magnitude increasing with the swirling strength. This increased back-scatter may be due to locally higher heat release rates in locations of flame/vortex interaction. Overall, the results demonstrate the importance of kinetic energy back-scatter at scales around the flame thickness and the complicated relationship between back-scatter and local flame/flow structure; these results should be considered when modeling turbulence in flames. 

\vspace{10pt}
\parbox{1.0\textwidth}{\footnotesize {\em Keywords:} Flame-Generated Turbulence; Subfilter-Scale Modeling; Turbulent Combustion; Laser Diagnostics; Swirl Flames}
\rule{\textwidth}{0.5pt}
\vspace{10pt}
\end{@twocolumnfalse}] 


\clearpage

\section{Introduction} \addvspace{10pt}
\label{Introduction}
Turbulent premixed combustion is characterized by non-linear and multi-scale interactions between chemistry, molecular transport, and fluid motion, leading to modification of both the flame structure and turbulence~\cite{Steinberg2020}. Coupling between chemical energy release and kinetic energy occurs due to thermal expansion (dilatation), typically at length scales associated with the flame (\textit{i.e.} the laminar flame thermal thickness $\delta_\mathrm{L}^0$). Increasing evidence \cite{Towery2016,OBrien2017,Kim2018, Kazbekov2021, MacArt2018} suggests that the presence of the flame can lead to qualitative deviations from the direct forward energy cascade of constant density turbulence at scales around $\delta_\mathrm{L}^0$, potentially resulting in mean up-scale kinetic energy transfer (back-scatter) within the flame. 

Using direct numerical simulations (DNS) of planar flames in homogeneous isotropic turbulence, Towery et al.~\cite{Towery2016} observed  reversal of advective energy transfer at scales around the laminar flame thickness. O'Brien \cite{OBrien2017} identified combustion-induced mean back-scatter when data were conditioned on the reaction progress variable $c$. Kim et al. \cite{Kim2018} reported an increase in the kinetic energy at small scales inside the flame, with the small scales becoming the donor of energy to larger scales. Experimentally, we recently quantified mean back-scatter in swirl-stabilized premixed flames across a range of Karlovitz numbers, up to $\mathrm{Ka} = 50$ \cite{Kazbekov2021}. These experiments confirmed the DNS observations and also demonstrated that mean back-scatter can occur in practical configurations. 

While our previous study investigated the influence of the local turbulence on cross-scale energy transfer via the local strain-rate field, we did not investigate the role of large-scale coherent flow structures; swirl flames often exhibit such structures, e.g. helical shear layer vortices~\cite{Sieber2017,Oberleithner2011,Steinberg2013,Terhaar2015}. These flow structures contain a significant portion of the fluctuating kinetic energy and are responsible for large-scale stirring (and hence flame corrugations, increased reaction rate, etc.). Consequently, there is a potentially complicated coupling between large-scale structures, heat release rate, and kinetic energy dynamics. Indeed, even without a flame, the presence of coherent structures may disrupt the local equilibrium between turbulence generation by the mean flow and dissipation at smaller scales~\cite{Holmes1996}.

In this paper, we extend our previous analysis to consider the simultaneous influence of large-scale flow structures and the flame on the kinetic energy cascade at scales around $\delta_\mathrm{L}^0$ in a premixed swirl flame. Following the physical-space analysis of O'Brien et al.~\cite{OBrien2017}, the transport equation for the kinetic energy of the filtered flow ($k \equiv \frac{1}{2} \Tilde{u}_i \Tilde{u}_i$) is 
\begin{equation}
\frac{\partial k}{\partial t} + \Tilde{u}_i \frac{\partial k}{\partial x_i} =
\underbrace{-\frac{\Tilde{u}_i}{\Bar{\rho}} \frac{\partial \Bar{p}}{\partial x_i}}_{\alpha_\mathrm{p}} + 
\underbrace{\frac{\Tilde{u}_i}{\Bar{\rho}} \frac{\partial \Bar{\tau}_{ij}}{\partial x_j}}_{\alpha_\nu}
\underbrace{-\frac{\Tilde{u}_i}{\Bar{\rho}} \frac{\partial \mathcal{T}_{ij}}{\partial x_j}}_{{\asfs}},
\label{e:k}
\end{equation}
where $\alpha_\mathrm{p}$, $\alpha_\nu$, and {\asfs} represent the contributions to resolved kinetic energy through work by resolved pressure, resolved viscous stresses, and sub-filter-scale stresses, respectively. The rate of kinetic energy transfer across the filter scale is identified as 
\begin{equation}
    {\asfs} = -\frac{\Tilde{u}_i}{\Bar{\rho}} \frac{\partial \mathcal{T}_{ij}}{\partial x_j}.
    \label{e:asfs}
\end{equation}
${\asfs}>0$ represents the local transfer of kinetic energy from scales smaller than $\Delta$ to scales larger than $\Delta$ (\textit{i.e.} back-scatter); downscale energy transfer (\textit{i.e.} forward-scatter) is indicated by ${\asfs}<0$. {\asfs} also appears in the equation for the sub-filter scale (SFS) kinetic energy ($k_\mathrm{sfs} \equiv \frac{1}{2}(\widetilde{u_i u_i} - \Tilde{u}_i \Tilde{u}_i)$) but with the opposite sign, demonstrating the two-way coupling between the resolved and SFS kinetic energies~\cite{OBrien2017}.

Hence, this study experimentally evaluates the simultaneous effects of heat release and large-scale vortex structures on ${\asfs}$ in a turbulent premixed swirl flame using tomographic particle image velocimetry (TPIV) and formaldehyde (\chem{CH_2O}) planar laser induced fluorescence (PLIF). Large-scale vortices are identified using the {{\lci}}-criterion~\cite{Zhou1999} and proper orthogonal decomposition (POD) is used to isolate a coherent periodic vortex. We then analyze the conditional statistics of {\asfs}, {\lci}, and the flame, both on an instantaneous basis and resolved with respect to the periodic vortex motion. We find that vortex-flame interactions lead to increased back-scatter compared to other regions of the flame.

\section{Experimental Setup} 
\begin{figure}[!t]
    \centering
    \subfigure[Swirl burner with with measurement locations. The inset shows the mean flow streamlines overlapping the mean progress-variable field at $z=0$~mm.]{\includegraphics[width=50 mm]{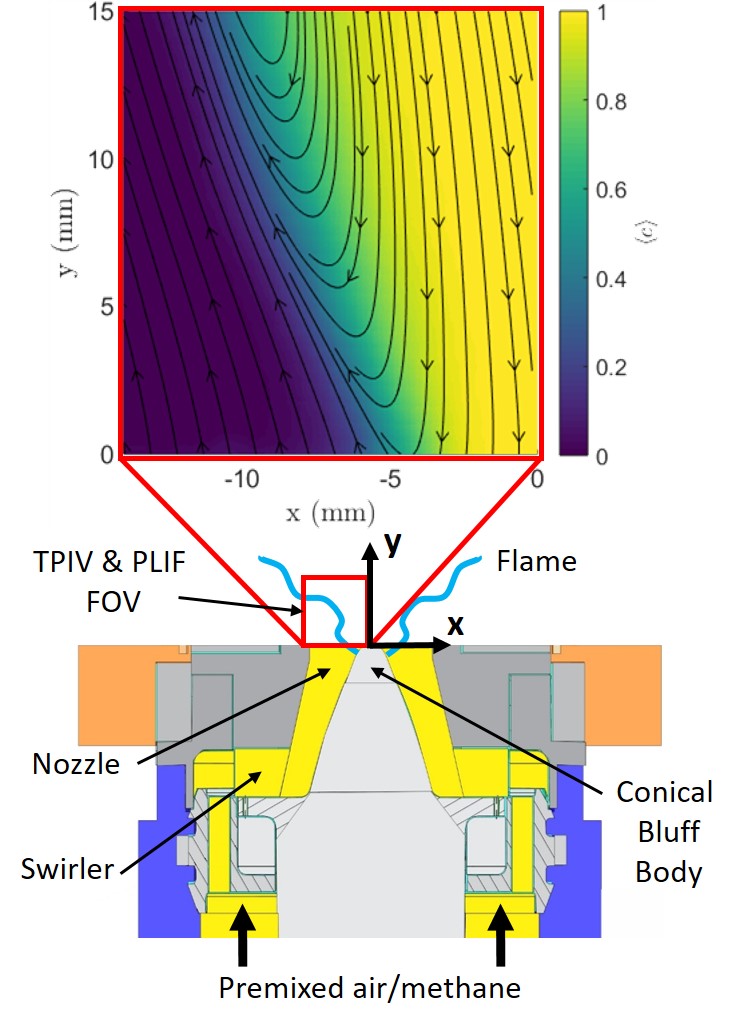}}\\
    \subfigure[Diagnostics configuration]{\includegraphics[trim = {30 0 30 0},clip,width= 60 mm]{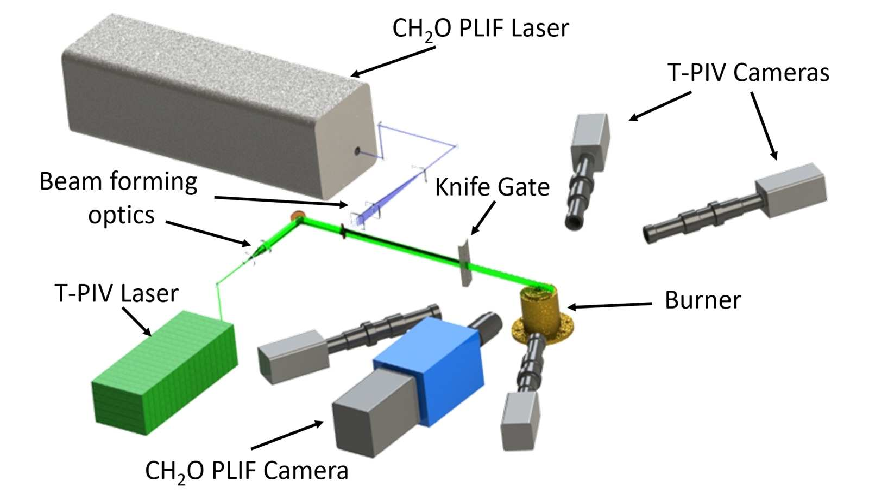}}
    \caption{Experimental configuration.}
    \label{f:setup}
\end{figure}

\addvspace{10pt}
The experimental setup is identical to that described in Refs.~\cite{Kazbekov2018,Kazbekov2020,Kazbekov2021} and only a brief summary is given here. The model gas turbine combustor (Fig.~\ref{f:setup}(a)) was idential to that originally described by Meier et al.~\cite{Meier2007}, but was operated without the combustion chamber to prevent window contamination from flow tracer particles (see below). The combustor features a 12-vane radial swirler and a 27.85~mm diameter nozzle with a conical bluff body along the centerline. The measured swirl number is $S=0.55$ at the exit plane of the nozzle. Flow rates of ambient temperature methane and air were controlled using electromechanical mass flow controllers (Brooks, 1\% full-scale uncertainty) and premixed far upstream of the combustor plenum.

Two test cases are considered here: a premixed flame at an equivalence ratio of $\phi=0.85$ and bulk flow rate of $U=25$~m/s, and a non-reacting air flow at the same bulk flow rate ($U$ is the ratio of the total volumetric flow rate and the nozzle exit area). The corresponding turbulence Reynolds number, Karloviz number, and Damk\"ohler number were $\mathrm{Re_T}=u^\prime\ell/\nu_\mathrm{r} = 3300$, $\mathrm{Ka} = (u^\prime/s_\mathrm{L}^0)^{3/2}(\delta_\mathrm{L}^0/\ell)^{1/2} = 50$, and $\mathrm{Da}=\big(\mathrm{Re_T}/\mathrm{Ka}^2 \big)^{1/2} = 1.1$. Here, $\nu_\mathrm{r}$ is the kinematic viscosity in the reactants, $u^\prime$ is the root-mean-squared velocity fluctuations in the shear layer between the inflowing reactants and central recirculation zone for the non-reacting flow and $\ell$ is the integral length scale, which was taken to be the full-width-half-maximum thickness of the shear layer. The laminar flame speed $s_\mathrm{L}^0$ and flame thickness $\delta_\mathrm{L}^0$, were calculated using the freely propagating flame model in Cantera with GRI3.0 chemical mechanism~\cite{Cantera}.

Measurements were made of the 3D velocity field and planar \chem{CH_2O} distribution using simultaneous TPIV and PLIF, respectively. The TPIV measurement region was a $16 \times 16 \times 1.3$~mm rectangular volume located directly above the nozzle, focused on the one branch of the axisymmetric swirl flame (see Fig.~\ref{f:setup}(a)). The PLIF field-of-view was along the centerline plane of the TPIV volume. The diagnostics configuration is shown in Fig.~\ref{f:setup}(b). A total of 1,000 simultaneous TPIV and PLIF measurements were made at each test condition.

For the TPIV, a series of lenses and two knife blades were used to shape the output of a dual-head Nd:YAG laser (Quantel Evergreen 200, 532~nm, 10~Hz, 200~mJ/pulse, $3~\mu$s between pulses) into a $1.3$~mm thick collimated rectangular slab. The air flow was seeded with alumina particles having a nominal diameter of $0.3$~\textmu{m} prior to mixing with the methane. The Stokes number of particles – based on a conservative estimate of the Kolmogorov time scale – was below $0.03$. 
 
 Mie scattering from the seed particles was captured by four sCMOS cameras (Andor Zyla 5.5, $2048\times2048$ pixels, 6.5~\textmu{m} pixel pitch), which were positioned on either side of the laser sheet at angles between $20^\circ$ and $30^\circ$ relative to laser propagation direction. A long-distance microscope (Infinity K2, CF-1/B objectives, $f/ \# =38$) was mounted to each camera via a Scheimpflug adapter (LaVision) to enable off-axis imaging. 
 
 Commercial software (LaVision DaVis 8.4) was used to reconstruct the particle tomograms and compute the 3D velocity vectors using direct spatial correlation. The final velocity resolution was 250~\textmu{m} with a 50\% vector overlap (125~\textmu{m} vector spacing). As described in our previous work~\cite{Kazbekov2018}, significant effort was made to optimize the spatial resolution and accuracy of the TPIV diagnostics; the results presented are converged with respect to the interrogation box size, indicating that the measurements are well resolved.

The \chem{CH_2O} PLIF system consisted of a sCMOS camera (Andor NEO 5.5, $2048\times2048$ pixels, 6.5~\textmu{m} pixel pitch) with a macro lens (Tamron, $f=180$~mm, $f/ \# =2.8$), image intensifier (LaVision IRO, gate = 100~ns), and an Nd:YAG laser (Quanta-Ray INDI-40-10, 355~nm, 10 Hz, 70~mJ/pulse). The laser sheet was formed using a set of cylindrical lenses and transmitted along the center of the TPIV volume (\textit{i.e.} $z=0$~mm plane). The laser excited the $A-X_0^1$ transition of \chem{CH_2O} around $355$~nm; fluoresence in the range of $370-480$~nm was isolated using a specialized filter (Semrock FF01-CH2O-50.8-D).

\section{Processing and data analysis} \addvspace{10pt}
\subsection{Determination of \asfs} \addvspace{10pt}
Determination of {\asfs} in this work uses the same methodology as our previous paper, which also includes a detailed assessment of the robustness and uncertainty of the method~\cite{Kazbekov2021}. While there are several approximations necessary in our empirical {\asfs} calculation, this previous study demonstrated that the results are quantitatively robust to these approximations. A brief summary of the methodology is provided here for completeness.

Calculation of {\asfs} requires filtering the instantaneous velocity and density fields. The 3D velocity fields are provided by the TPIV and we estimate the 2D density field from the \chem{CH_2O} PLIF. In methane/air flames, \chem{CH_2O} is formed at early stages of fuel breakdown in a relatively inert ``preaheat zone'' and rapidly consumed during \chem{HCO} formation in the thin region of rapid exothermic reactions; \chem{CH_2O} serves as a good indicator of the instantaneous thermal width $\delta_\mathrm{L}^0$. Here, we use the \chem{CH_2O}-containing regions to identify the instantaneous flame boundaries. We then define a progress-variable-like parameter $c=(d-d_r)/(d_p-d_r)$, where $d_r$ and $d_p$ are the shortest in-plane distances to the reactant and product facing boundaries of the \chem{CH_2O}-containing region; our $c$ maps the distance of a location through the instantaneous flame surface. 

The local density and temperature are then approximated to be the same as a laminar flame with the same reactants at the corresponding $c$. We highlight and acknowledge that the \chem{CH_2O}-to-$\rho$ mapping is not exact. However, we have demonstrated that the calculation of {\asfs} -- the primary quantity of interest -- does not depend on the detailed density distribution due to the spatial filtering inherent in the analysis \cite{Kazbekov2021}. Indeed, even a simple bimodal density approximation (jumping between the density of the reactants and products) provides qualitatively equivalent results. 

Calculating {\asfs} via Eq.~\ref{e:asfs} requires filtered velocity and density fields, along with their spatial derivatives. We consider a top-hat filter with $\Delta/\delta_\mathrm{L}^0 \approx 1.5$ ($\Delta = 6$ vectors), capturing the cross-scale transfer of kinetic energy between the flame scales and the larger scales. While the 3D measured velocity fields enable computation of the required derivatives, density is only available in the central TPIV plane. Hence, we restrict our analysis to this plane and use conservation of mass to approximate the out-of-plane density gradient needed for computation of {\asfs} as
\begin{multline}
\frac{\partial \bar{\rho}}{\partial x_3} = \frac{1}{\Tilde{u}_3} \left[-\bar{\rho} \left(\frac{\partial \Tilde{u}_1}{\partial x_1} + \frac{\partial \Tilde{u}_2}{\partial x_2} + \frac{\partial \Tilde{u}_3}{\partial x_3} \right)\right.\\ - \left.\Tilde{u}_1 \frac{\partial \bar{\rho}}{\partial x_1} - \Tilde{u}_2 \frac{\partial \bar{\rho}}{\partial x_2} - \frac{\partial \bar{\rho}}{\partial t}\right]
\label{eq:drho_dz}
\end{multline}
We present our results in terms of the ensemble averaged {\asfs} (denoted $\langle \asfs \rangle$) and hence the contribution from the unsteady density term is negligible in the statistically stationary turbulent flow studied here.

\subsection{Identification of coherent structures} \addvspace{10pt}
We use the {\lci}-criterion to identify coherent swirling eddies in this variable density flow; eddies are spatial regions in which the velocity gradient tensor has a complex eigenvalue pair ($\lambda_\mathrm{cr} \pm i{\lci}$) \cite{Zhong1998,Kolar2007}. The imaginary part of the complex eigenvalue pair - {\lci} - quantifies the strength of the local swirling motion. Physically, the local flow is either stretched or compressed along the axis of the real eigenvector ($\boldsymbol{\nu_r}$), while the flow is swirling in the plane spanned by the real and imaginary parts of the complex eigenvector ($\boldsymbol{\nu_{cr}}$ and $\boldsymbol{\nu_{ci}}$). 

We use {\lci} fields to identify the position and size of the largest scale coherent structures in the flow. Hence, for the {\lci} calculation, we first spatially smooth the velocity fields using a 21 vector box filter. This smoothed velocity field retains large scale eddies while suppressing smaller structures. To ensure that this smoothing does not affect our conclusions, we re-performed the analysis using 11-vector  and 6-vector filters. While the magnitude of peak {\lci} decreased with increasing filter size, the largest eddies retained their size and shape.

\begin{figure}[!t]
    \includegraphics[width=192pt]{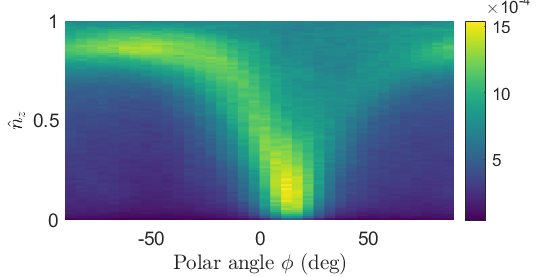}
    \caption{Joint probability density function of $\boldsymbol{\Hat{n}_z}$ and the polar angle $\phi$ of the normal vector in the $x$-$y$ plane.}
    \label{f:pdf_vecnorm}
\end{figure}

The eigenvalues and eigenvectors were computed on the central plane of the TPIV volume. Locations with three real eigenvalues were assigned a value of ${\lci}=0$ (\textit{i.e.} no swirl); the magnitude of {\lci} was retained at all other locations to quantify the swirl. The data were further reduced by eliminating locations where the axis of rotation was predominantly in the $x$-$y$ plane. To demonstrate the distribution of rotation axis orientations, we consider the normal vector to the plane of rotation, defined as $\boldsymbol{\Hat{n}}=\boldsymbol{\nu_\mathrm{cr}} \times \boldsymbol{\nu_\mathrm{ci}}$. Figure~\ref{f:pdf_vecnorm} shows the joint probablity density function (PDF) of $\boldsymbol{\Hat{n}_z}$ and the polar angle $\phi$ of $\boldsymbol{\Hat{n}}$ in the $x$-$y$ plane. The locations with $\phi \approx 0-25$~degrees were found to correspond to the large scale swirl induced by the radial swirl vanes. However, the locations with large $\boldsymbol{\Hat{n}_z}$ correspond to coherent structures that rotate in plane. We therefore set ${\lci}=0$ at locations with $\boldsymbol{\Hat{n}_z} < 1/\sqrt{2}$ to isolate the structures of interest.

\subsection{Proper orthogonal decomposition} \addvspace{10pt}

\begin{figure}[t]
    \centering
    \subfigure[Mode 1.]{\includegraphics[width=0.48\columnwidth]{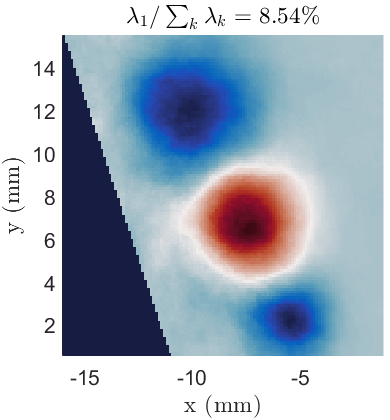}} \hfill
    \subfigure[Mode 2.]{\includegraphics[width=0.48\columnwidth]{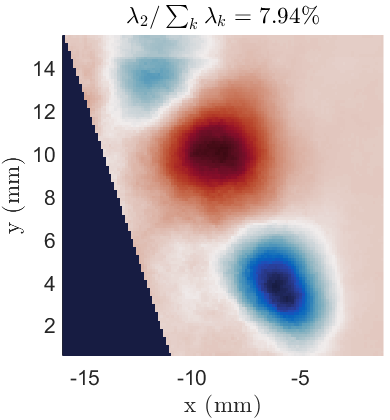}}
    \caption{First two spatial POD modes for the reacting flow.}
    \label{f:modes12}
\end{figure}

\begin{figure*}
    \includegraphics[width=143mm]{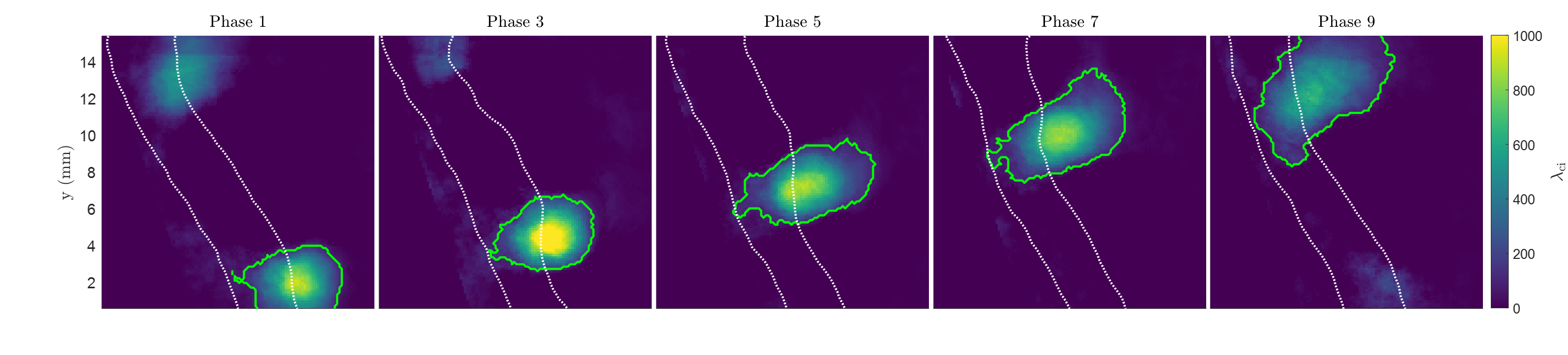}
    \caption{Phase averaged evolution of the dominant coherent vortex $\langle {\lci} | \theta_i \rangle$, with every second phase shown. White dashed curve represents $\langle c | \theta_i \rangle = 0.1$ and $0.9$. Solid green curve is an isocontour of $\langle {\lci} | \theta_i \rangle = \frac{1}{8}\langle {\lci} | \theta_i \rangle_\mathrm{max}$.}
    \label{f:modes}
\end{figure*}

We use the method of snapshots POD~\cite{Sirovich1987} to extract the dominant coherent structures, from which we phase-sort individual measurements based on phase angle of the structure motion. Given a set of $N$ measurements of the variable $\xi(\boldsymbol{x},t)$, the fluctuating component ($\xi'=\xi - \langle \xi \rangle$) can be approximated using $P$ modes as

\begin{equation}
    \xi'(\boldsymbol{x},t) = \sum_{j=1}^P a_j(t) \phi_j(\boldsymbol{x}),
\end{equation}
where $\phi_j(\boldsymbol{x})$ and $a_j(t)$ represent the spatial modes and their temporal coefficients, respectively. POD modes $\phi_j$ are given by the eigenvectors of the covariance matrix $\boldsymbol{R} = XX^T$, where $X$ represents the $N$ snapshots in a matrix form. The eigenvalues ($\lambda_j$) are the mode energies, which describe how much of the variance in the original data is contained in a particular mode.

We use the measured 2D swirling strength fields on the center plane of the TPIV volume as the variable $\xi(\boldsymbol{x},t)$ for POD. Figure~\ref{f:modes12} shows the first two spatial modes for the reacting flow, which account for about 16.5\% of variance. Both modes clearly show structures resembling vortices propagating along the inner shear layer; mode 2 is translated by approximately a quarter of the wavelength. Moreover, the modes have similar relative energy, suggesting the pair forms a single vortex that appears to convect downstream.  

Previously, the dominant dynamic flow structure in this burner was identified as either a helical vortex core or a toroidal vortex, depending on the operating conditions \cite{Steinberg2013}. The structure identified here corresponds to one of these shear layer vortex modes, though the exact mode cannot be identified due to the field of view, which only covers one half of the burner. Based on the operating conditions and results in Ref.~\cite{Steinberg2013}, we expect this to be a helical mode. The POD on the non-reacting test case showed a similar set of modes and relative energies as the reacting flow, again consistent with the helical modes observed in all non-reacting cases in Ref.~\cite{Steinberg2013}. Regardless, the exact nature of the flow structure is not critical for the analysis presented here.

We use time coefficients for modes 1 and 2 to obtain phase information of the periodic flow structure. This is done by phase-sorting individual measurements into nine equal ``bins'' based on the angle of the measurement in the $a_1(t)$ and $a_2(t)$ plane, e.g.~\cite{Gilge2020,Oudheusden2005,Perrin2007,Legrand2011}. Phase-averaged quantities of interest $q$ are calculated as
\begin{equation}
    \langle q|\theta_i \rangle = \frac{1}{N_{\theta_i}} \sum_{k=1}^{N_{\theta_i}} q(t_k),
    \label{e:ph_avg}
\end{equation}
where $\theta_i$ is the $i$-th range of phase angles and $N_{\theta_i}$ is the ensemble size. We consider {\lci}, \asfs, and $c$ as the primary quantities of interest in the analysis below. 

Figure~\ref{f:modes} shows the profiles of $\langle {\lci} | \theta_i \rangle$ at different phases of propagation in a reacting flow. Every second phase is shown for brevity and the solid lines denote the phase-conditioned flame brush, i.e. $0.1<\langle c|\theta_i\rangle<0.9$. The vortex moves downstream along the inner shear layer between the flame and the recirculating products. The size of the vortex grows and the peak {\lci}~decreases due to diffusion and flow dilatation as the vortex moves downstream. The structure and shape of the flame brush also changes in response to the presence of the vortex.

\section{Results and Discussion} \addvspace{10pt}
\subsection{Instantaneous Data and Conditional Statistics} \addvspace{10pt}

\begin{figure*}[!h]
    \centering
    \subfigure[{\lci}]{\includegraphics[width=70 mm]{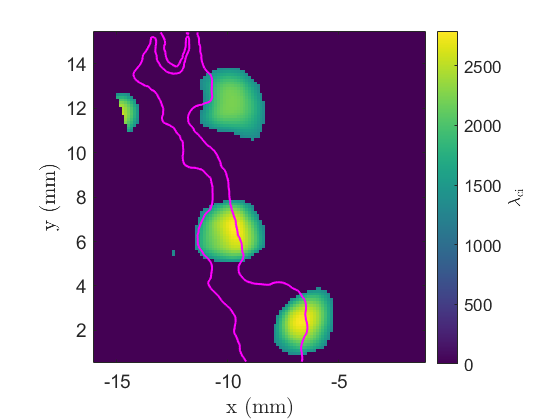}}
    \subfigure[\asfs]{\includegraphics[width=70 mm]{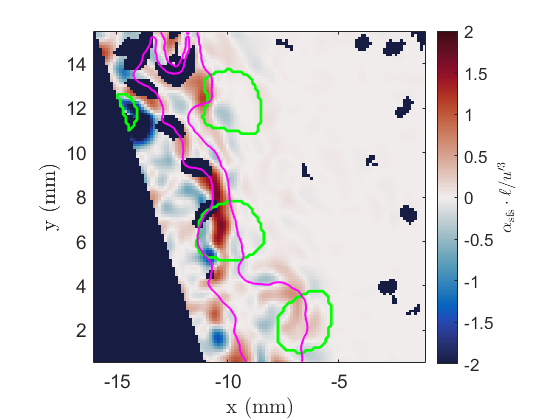}}
    \caption{Instantaneous realization of {\lci}~and {\asfs} for $\Delta/\delta_\mathrm{L}^0=1.5$. Magenta curves correspond to $c=0.1$ and $c=0.9$ contours. Green curve is an isocontour of ${\lci}=\frac{1}{2}\max({\lci})$ at this instant.}
    \label{f:single_snap}
\end{figure*}

We first examine the relationship between {\lci}, the flame, and {\asfs} on an instantaneous basis. Figure~\ref{f:single_snap}(a) shows a representative snapshot of {\lci}, overlaid with $c=0.1$ and $c=0.9$ contours as markers of the flame. A set of large-scale coherent swirling structures (eddies) intersects with the instantaneous flame. The flame generally is thicker and more curved at locations of flame/vortex interaction.

Figure~\ref{f:single_snap}(b) shows the corresponding {\asfs} field with contours of the flame and eddies shown in magenta and green, respectively. Energy transfer occurs in both the up-scale (back-scatter, red) and down-scale (forward-scatter, blue) directions across the filter scale $\Delta$, with regions of energy back-scatter predominantly occur in the vicinity of the flame. The magnitude of {\asfs} is substantially lower in the products due to suppression of small-scale motion as a result of flow dilatation and increasing viscosity across the flame. 

\begin{figure*}[!t]
    \centering
    \subfigure[PDFs of {\asfs} conditioned on {$1000<\lci<2000$}.]{\includegraphics[width=70 mm]{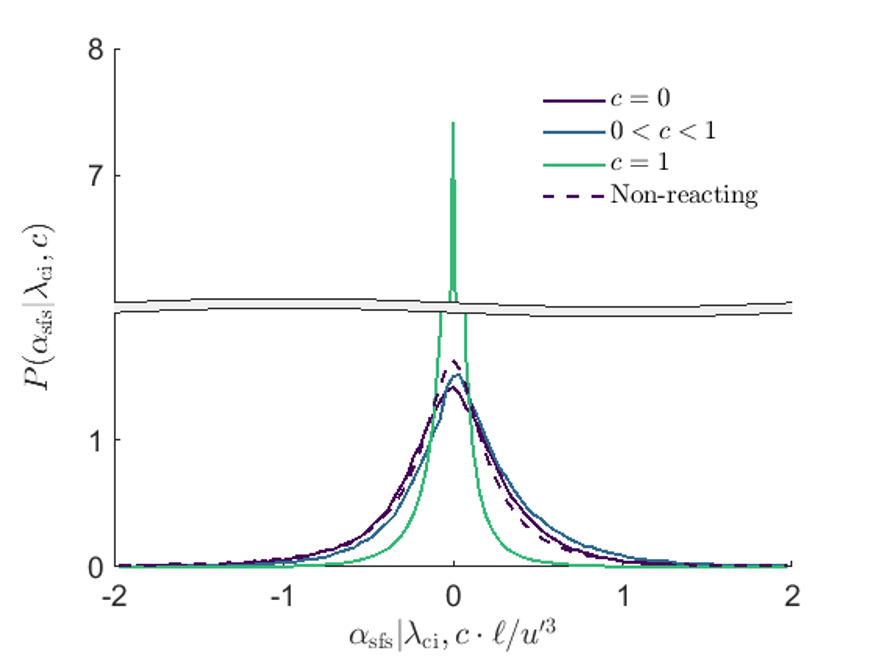}}
    \subfigure[Mean of $P(\asfs|{\lci})$ internal and external to the flame.]{\includegraphics[width=70 mm]{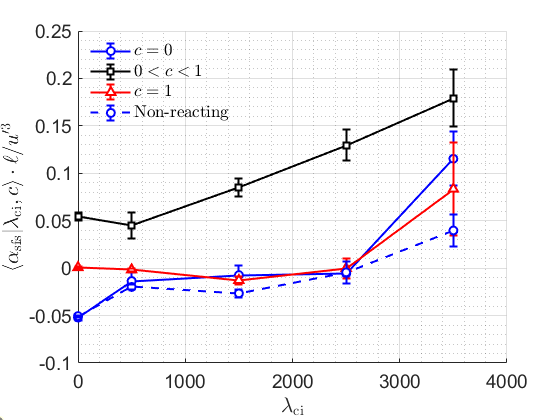}}
    \caption{Variation of instantaneous {\asfs} with {\lci}~ for reacting and non-reacting flows.}
    \label{f:single_snap_cond}
    \vspace{-1em}
\end{figure*}

Since it is difficult to ascertain correspondence between the eddies and cross-scale energy transfer from this individual image, the conditional PDFs of {\asfs} are plotted in Fig.~\ref{f:single_snap_cond}(a); the PDFs are conditioned on moderate swirl ($1000<\lci<2000$) and the reacting data are conditioned on the pure reactants ($c=0$), internal to the flame ($0<c<1$) and in the burnt products ($c = 1$). In the reactants and in non-reacting flow, the PDFs are nearly symmetric, indicating nearly equal forward- and back-scatter across $\Delta$. Similarly, the PDF in the products is nearly symmetric, but with reduced range of {\asfs} due to attenuation of turbulence at elevated temperature. Inside the flame, the PDF is positively skewed, indicating an increased probability of back-scatter.

To further articulate the simultaneous effects of the flame and eddies on {\asfs}, Fig. ~\ref{f:single_snap_cond}(b) shows the conditional mean $\langle\asfs|{\lci}, c\rangle$ for both the reacting and non-reacting flows (the PDFs in Fig.~\ref{f:single_snap_cond}(a) are from the data comprising the mean at $\lci = 1500$ in Fig. ~\ref{f:single_snap_cond}(b)). The error bars indicate the expected uncertainty in the mean based on the sample standard deviation and number of samples, approximating the statistics as normal. For the non-reacting case, $\langle\asfs\rangle$ is negative for most values of {\lci}, obtaining a slightly positive value at locations of high swirl. The data in the reactants and products of the reacting case are qualitatively similar to the non-reacting case. Most interesting are the data within the flame, which clearly demonstrate the flame-induced mean back-scatter and that the back-scatter magnitude increases with the local swirling strength. Regions of locally high swirling strength are expected to correspond to local increases in reaction rate due to flame wrinkling and increased scalar gradients. Hence, the data are indicative of a positive correlation between local heat release rate and back-scatter.

\subsection{Phase-Conditioned Analysis} \addvspace{10pt}
The above results demonstrate how the presence of swirling eddies and flame affect the local cross-scale kinetic energy transfer. We now isolate the impact of the single most dominant coherent structure on the energy dynamics, namely the periodic vortex identified from the POD (see Fig.~\ref{f:modes}). To do so, we examine the phase-averaged {\lci}, {\asfs}, and $c$ profiles generated using POD modes 1 and 2 in the reacting and non-reacting flows.

\begin{figure}[!t]
\centering
    \includegraphics[width=0.9\columnwidth]{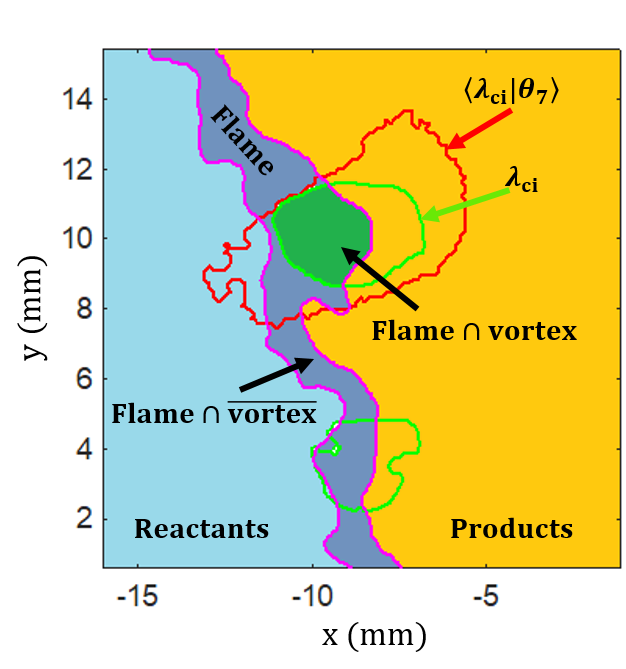}
    \caption{Schematic of the regions used for conditioning the phase-averaged statistics, taken from an instantaneous image at phase 7.}
    \label{f:regions}
\end{figure}

\begin{figure}[!t]
    \subfigure[$\langle \asfs | \theta_i, \mathrm{region} \rangle$]{\includegraphics[width=\columnwidth]{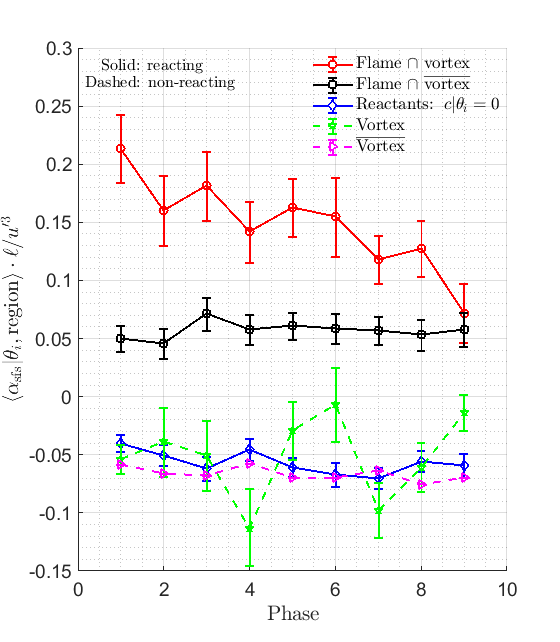}}
    \subfigure[$\langle \lci | \theta_i, \mathrm{region} \rangle$]{\includegraphics[width=\columnwidth]{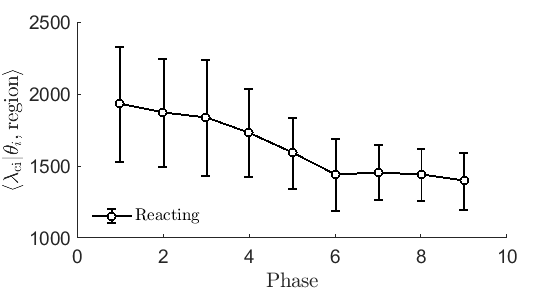}}
    \caption{Phase-averaged statistics as a function of phase for the most dominant coherent structure in the reacting and non-reacting flows.}
    \label{f:modes_avg}
    \vspace{-1em}
\end{figure}

Phase averaging is performed based on Eq.~\ref{e:ph_avg}, and we further condition on specific spatial regions of interest. To demonstrate this, Fig.~\ref{f:regions} shows an instantaneous schematic of {\lci} and $c$ in phase 7, showing the location of the phase-averaged vortex ($\langle \lci | \theta_i \rangle$ - red contour), the location of the instantaneous vortex ({\lci} - green contour), and the region occupied by the flame ($0<c<1$ - magenta contour). The ``$\mathrm{flame}~\cap~\mathrm{vortex}$'' region (shaded green) represents the region where the flame, {\lci}, and $\langle \lci | \theta_i \rangle$ intersect; the instantaneous vortex that lies outside the $\langle \lci | \theta_i \rangle$ contour is not part of the periodic vortex and, hence, is not considered in this analysis. The ``$\mathrm{flame}~\cap~\overline{\mathrm{vortex}}$'' region (shaded purple) represents the locations inside the flame but external to the instantaneous vortex inside the $\langle \lci | \theta_i \rangle$ contour. We also consider ``$\mathrm{vortex}$'' and ``$\overline{\mathrm{vortex}}$'' which represent overlapping regions of {\lci} and $\langle \lci | \theta_i \rangle$ and its complement set, respectively. Finally, the ``$\mathrm{reactants}$'' region (shaded light blue) represents regions of $c=0$. We do not present data in the products (shaded orange), as {\asfs} generally is low in this region due to the attenuated turbulence.

Figure~\ref{f:modes_avg}(a) shows the phase-averaged {\asfs} in different regions of the flow as a function of phase. In the reactants and in the non-reacting flow, $\langle \asfs | \theta_i\rangle$ is negative and no significant phase dependence is observed. For the non-reacting flow, the increased scatter for regions containing the periodic vortex likely is attributed to uncertainty and convergence. Broadly speaking, the non-reacting flow and non-reacting regions of the reacting flow both show mean forward-scatter across the filter scale. 

The regions containing the flame show qualitative differences in {\asfs} compared to the non-reacting flows, with further qualitative differences between regions with and without the dominant periodic vortex. In the ``$\mathrm{flame}~\cap~\overline{\mathrm{vortex}}$'' region, $\langle \asfs | \theta_i, \mathrm{flame}~\cap~\overline{\mathrm{vortex}}\rangle$ is positive. This is consistent with our previous experimental findings~\cite{Kazbekov2021} and other DNS studies~\cite{Towery2016,OBrien2017,Kim2018}, demonstrating that the flame leads to mean back-scatter at scales around the laminar flame thickness. Importantly, regions where the flame and periodic vortex intersect exhibit higher mean back-scatter compared to regions of the flame without the vortex present, \textit{i.e.}  $\langle \asfs | \theta_i, \mathrm{flame}~\cap~{\mathrm{vortex}}\rangle> \langle \asfs | \theta_i, \mathrm{flame}~\cap~\overline{\mathrm{vortex}}\rangle$ . 

The magnitude of $\langle \asfs | \theta_i, \mathrm{flame}~\cap~{\mathrm{vortex}} \rangle$ decreases with phase. Figure~\ref{f:modes_avg}(b) shows that the magnitude of the swirl, \textit{i.e.} $\langle {\lci} | \theta_i, \mathrm{flame}~\cap~\mathrm{vortex} \rangle$, also decreases with phase as the vortex moves downstream. Hence, there is a correspondence between the swirling strength of the dominant periodic vortex and the mean back-scatter magnitude at locations of flame/vortex interaction. This result is consistent with the instantaneous data in Fig.~\ref{f:single_snap_cond}(b) and further articulates the simultaneous and complimentary effects of combustion induced heat release rate and large-scale coherent vortices on kinetic energy back-scatter. 

\section{Conclusion} \addvspace{10pt}
This study has experimentally investigated the cross-scale energy transfer in turbulent premixed swirl flames in the presence of coherent structures using high-resolution TPIV and \chem{CH_2O} PLIF. Coherent structures were identified using the {\lci}-criterion and POD was performed on {\lci}-fields to isolate the most dominant periodic coherent vortex structure. Cross-scale energy transfer {\asfs} was computed across a filter scale $\Delta/\delta_\mathrm{L}^0=1.5$ and conditional statistics of {\asfs} on {\lci} and the flame were compiled. 

The main conclusions are that (i) the non-reacting case and non-reacting regions of the the reacting case (pure reactants and products) exhibited mean down-scale transfer of kinetic energy, \textit{i.e.} forward-scatter, with nearly symmetric {\asfs} PDFs about $\asfs = 0$; (ii) reacting regions of the flow (inside the instantaneous flame surface) exhibited mean up-scale transfer of kinetic energy, \textit{i.e.} back-scatter, with slightly positively skewed PDFs; (iii) the mean magnitude of the back-scatter inside the flame was increased in regions of flame/vortex interaction, as demonstrated both through the instantaneous data and statistics conditioned on the phase of the periodic vortex structure; and (iv) there was a correspondence between the magnitude of the local swirling strength and the mean back-scatter magnitude. It is possible that the observed increased back-scatter in regions of flame/vortex interaction is due to greater local heat release rates, though further measurements would be needed to confirm this. Regardless, this study articulates the complicated influences of the flame and flow/flame interactions on turbulence dynamics around the flame scale, which should be considered in turbulence modeling.

\section{Acknowledgments} \addvspace{10pt}
This work was supported by the US Air Force Office of Scientific Research under grant FA9550-17-0011, Project Monitor Dr. Chiping Li. A. Kazbekov acknowledges the support of the NSERC-PGS D fellowship.
\vspace{-0.2em}

 \footnotesize
 \baselineskip 9pt


\bibliographystyle{pci}
\bibliography{references.bib} 


\newpage

\small
\baselineskip 10pt



\end{document}